\documentclass{ws-p8-50x6-00}
\usepackage{amsmath}
\usepackage{graphicx}
\def\e6{$E(6)$}
\def\10{$SO(10)$}
\def\21{$SU(2) \otimes U(1) $}
\def\lr{$SU(2)_L \otimes SU(2)_R \otimes U(1)$ }
\def\422{$SU(4) \otimes SU(2) \otimes SU(2)$}
\def\321{$SU(3) \otimes SU(2) \otimes U(1)$}

\newcommand{\ed}{\end{document}}
\DeclareMathAlphabet{\mathsc}{OT1}{cmr}{m}{sc}

\def\lsim{\raise0.3ex\hbox{$\;<$\kern-0.75em\raise-1.1ex\hbox{$\sim\;$}}}
\def\gsim{\raise0.3ex\hbox{$\;>$\kern-0.75em\raise-1.1ex\hbox{$\sim\;$}}}

\newcommand{\CL}   {C.L.}
\newcommand{\dof}  {d.o.f.}

\newcommand{\eVq}  {\rm{eV}^2}
\newcommand{\Sol}  {\mathsc{sol}}
\newcommand{\Atm}  {\mathsc{atm}}

\newcommand{\Lsnd} {\mathsc{lsnd}}

\newcommand{\Dms}  {\Delta m^2_\Sol}
\newcommand{\Dma}  {\Delta m^2_\Atm}
\newcommand{\Dml}{\Delta m^2_\Lsnd}

\def \nbb {$\beta\beta_{0\nu}$ }

\newcommand\newblock{\hskip .11em}

\newcommand{\AddrAHEP}{%
 Instituto de F\'{\i}sica Corpuscular,
  C.S.I.C. -- Universitat de Val{\`e}ncia \\
  Edificio de Institutos de Paterna, Apartado 22085,
  E--46071 Val{\`e}ncia, Spain\\}

\def\[{\left [}
\def\]{\right ]}
\def\({\left (}
\def\){\right )}

\begin{document}  

\title{Brief Neutrino Physics Update}

\author{J. W. F. Valle}

\address{\AddrAHEP}

\maketitle

\abstracts{ The discovery of neutrino mass establishes the need for
  physics beyond the Standard Model. I summarize the status of two--
  and three--neutrino oscillation parameters from current solar,
  atmospheric, reactor and accelerator data.  Future neutrinoless
  double beta decay experiments will probe the nature of neutrinos, as
  well as the absolute scale of neutrino mass, also tested by tritium
  beta decay spectra and cosmological observations.  Sterile neutrinos
  do not provide a good way to account for the LSND hint, which needs
  further confirmation.  Finally I sketch the main theoretical ideas
  for generating neutrino mass.}

\section{Two--Neutrino Parameters}
\label{sec:1}

In conjunction with the most recent SNO data with enhanced neutral
current sensitivity (salt phase)~\cite{Ahmed:2003kj} and the KamLAND
reactor data~\cite{eguchi:2002dm}, solar neutrino experiments have now
established the oscillation phenomenon. This closes the solar neutrino
problem and opens an era of opportunity for learning more about the
Sun~\cite{burgess:2002we} or about beyond--oscillations properties of
neutrinos, such as magnetic moments~\cite{schechter:1981hw} and
non-standard
interactions~\cite{wolfenstein:1978ue,mikheev:1985gs,valle:1987gv}.
Although well-motivated by theory, such mechanisms can no longer
account for the data and may only be present at a sub--leading
level~\cite{barranco:2002te,guzzo:2001mi,grimus:2002vb}.  Similarly,
the solid oscillation interpretation of the atmospheric neutrino
data~\cite{fukuda:1998mi} leaves little room for beyond-oscillation
non-standard physics~\cite{fornengo:2001pm}.

\begin{figure}[t] \centering
\includegraphics[width=.7\linewidth,height=5.5cm]{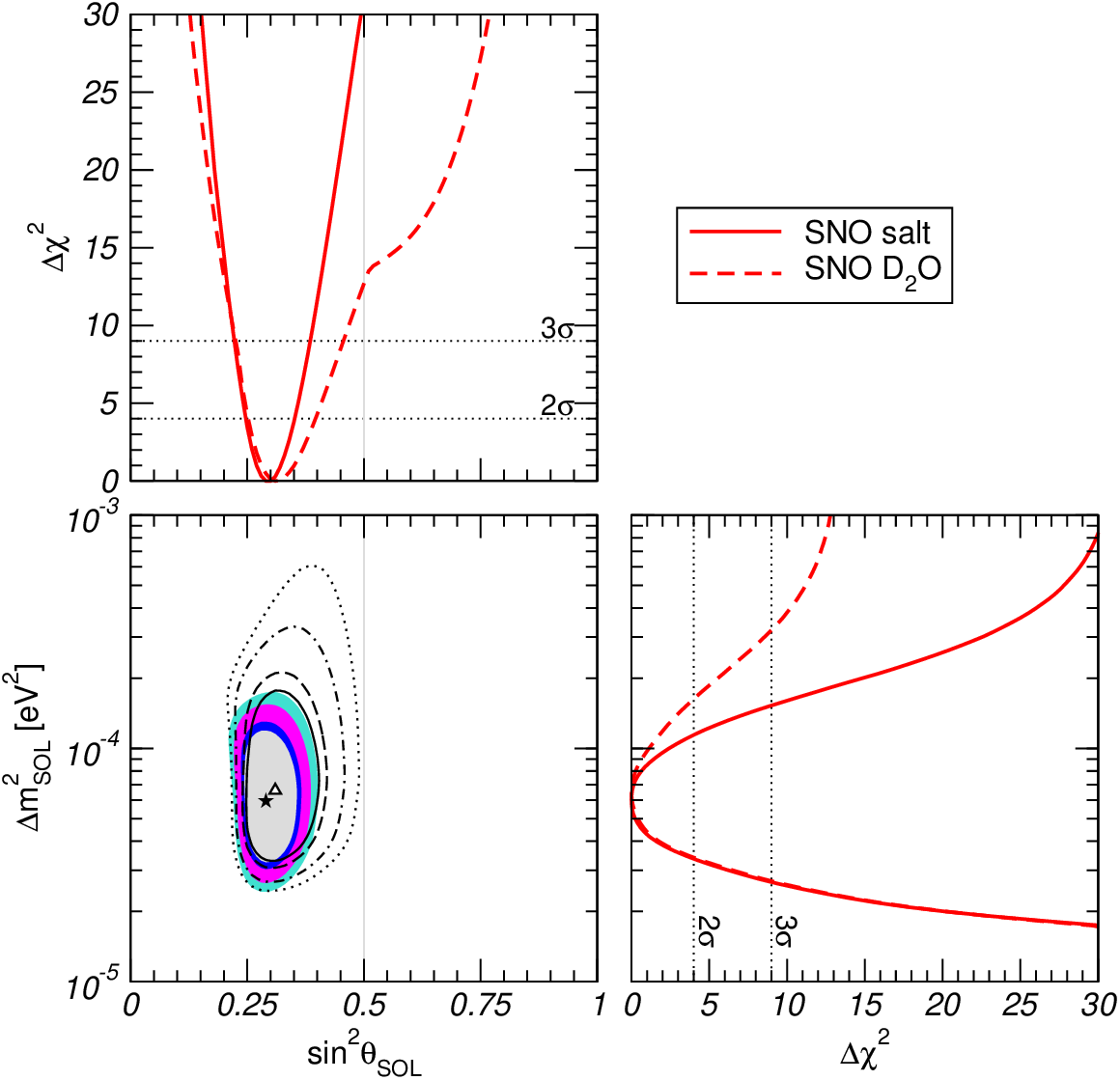}
  \caption{\label{fig:sol-region}%
    Two-neutrino solar neutrino oscillation parameters, from
    Ref.~\protect\cite{maltoni:2003da}. }  
\vglue-.5cm
\end{figure}

Neutrino
masses~\cite{gell-mann:1980vs,yanagida:1979,schechter:1980gr,mohapatra:1981yp,chikashige:1981ui,schechter:1982cv}
have finally been discovered~\cite{pakvasa:2003zv}.  A complete
analysis of recent solar, atmospheric, accelerator and reactor
neutrino data has been given in Ref.~\cite{maltoni:2003da}
\footnote{See Refs.~\cite{maltoni:2002ni,maltoni:2002aw} for extensive
  list of experimental solar and atmospheric neutrino references. For
  a discussion of other neutrino data analyses see Table 2 in
  \cite{maltoni:2002ni} and the reviews in~\cite{pakvasa:2003zv}.}.
This paper presents an updated determination of the neutrino
oscillation parameters taking into account all data (see
Ref.~\cite{maltoni:2003da} for details) including the new solar
neutrino data from the SNO--salt phase~\cite{Ahmed:2003kj}. The
resulting 90\%, 95\%, 99\%, and 3$\sigma$ 2 \dof\ \CL\ regions in
$\sin^2\theta_\Sol$, $\Dms$ allowed by all solar neutrino data before
(lines) and after (shaded regions) the inclusion of the SNO--salt data
are shown in Fig.~\ref{fig:sol-region}.  Also shown in this figure is
$\Delta \chi^2$ as a function of $\sin^2\theta_\Sol$ and $\Dms$,
minimized with respect to the undisplayed parameter.
One finds that especially the upper part of the LMA--MSW region and
large mixing angles are strongly constrained by the new data, with
$\sin^2\theta_\Sol = 0.5$ excluded at more than 5$\sigma$. This rules
out all bi--maximal models of neutrino mass~\cite{pakvasa:2003zv}.

\begin{figure}[h] \centering
\includegraphics[width=.7\linewidth,height=5.5cm]{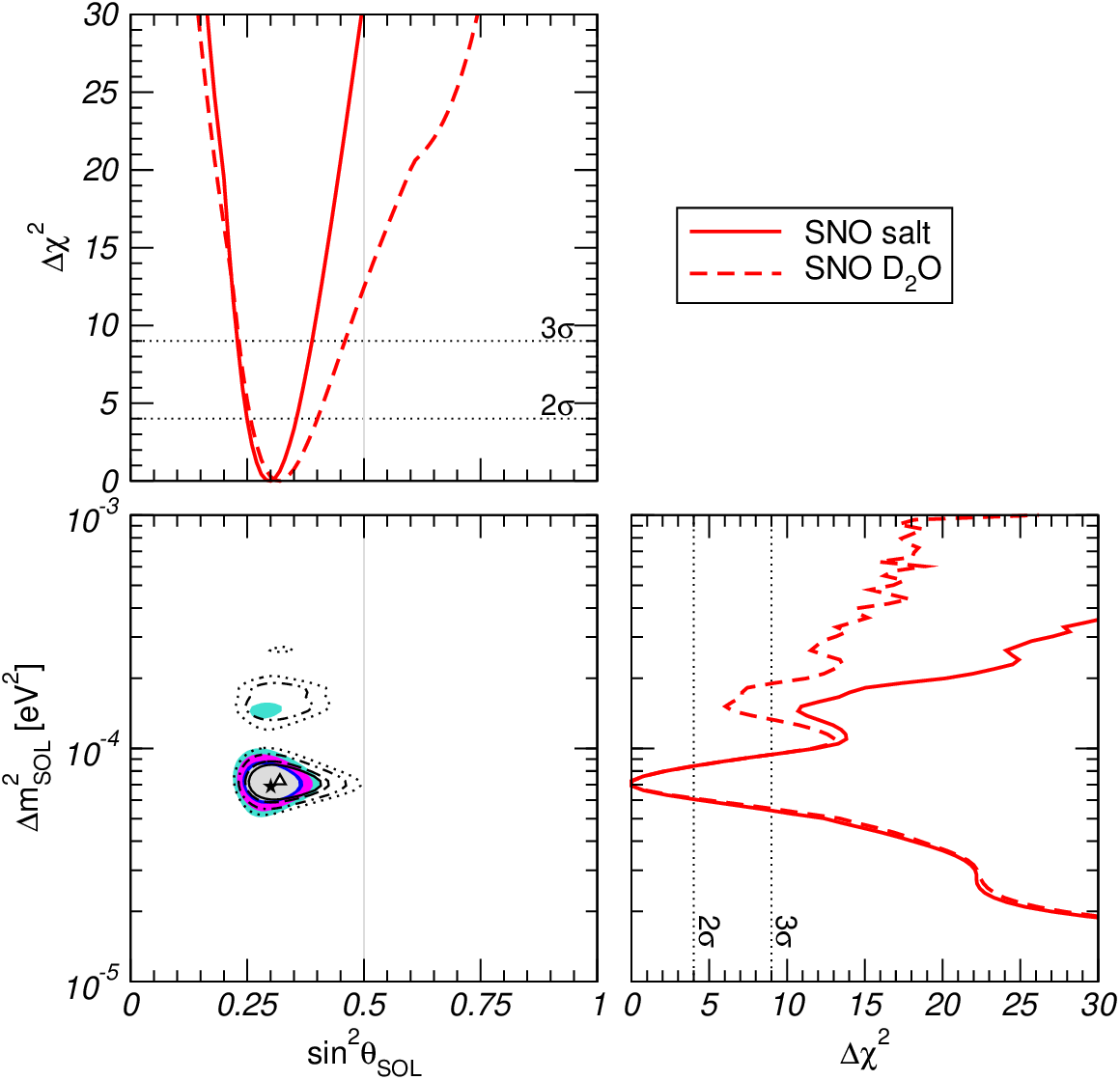}
    \caption{\label{fig:sol+kaml-region}%
      Two-neutrino solar+KamLAND neutrino oscillation parameters, from
      Ref.~\protect\cite{maltoni:2003da}.}  \vglue-.5cm
\end{figure}

The first 145.1 days of KamLAND data have important implications on
the determination of the solar neutrino parameters, as discussed,
for example, in Ref.~\cite{maltoni:2002ni,maltoni:2002aw}.
Fig.~\ref{fig:sol+kaml-region} shows the projections of the allowed
regions from all solar neutrino and KamLAND data at 90\%, 95\%, 99\%,
and 3$\sigma$ \CL\ for 2 \dof\ onto the plane of $\sin^2\theta_\Sol$
and $\Dms$ before (lines) and after (shaded regions) the inclusion of
the SNO--salt data.  Also shown is $\Delta \chi^2$ as a function of
$\sin^2\theta_\Sol$ and $\Dms$, minimized with respect to the
undisplayed parameter.
One sees that the SNO--salt results reject the previously allowed
high-mass branch of $\Dms$ at about 3$\sigma$.  Moreover, for the
first time it is possible to obtain meaningful bounds on solar
neutrino parameters at the 5$\sigma$ level, showing that neutrino
physics has just entered the precision age.

\begin{figure}[t] \centering
\includegraphics[width=.6\linewidth,height=5.5cm]{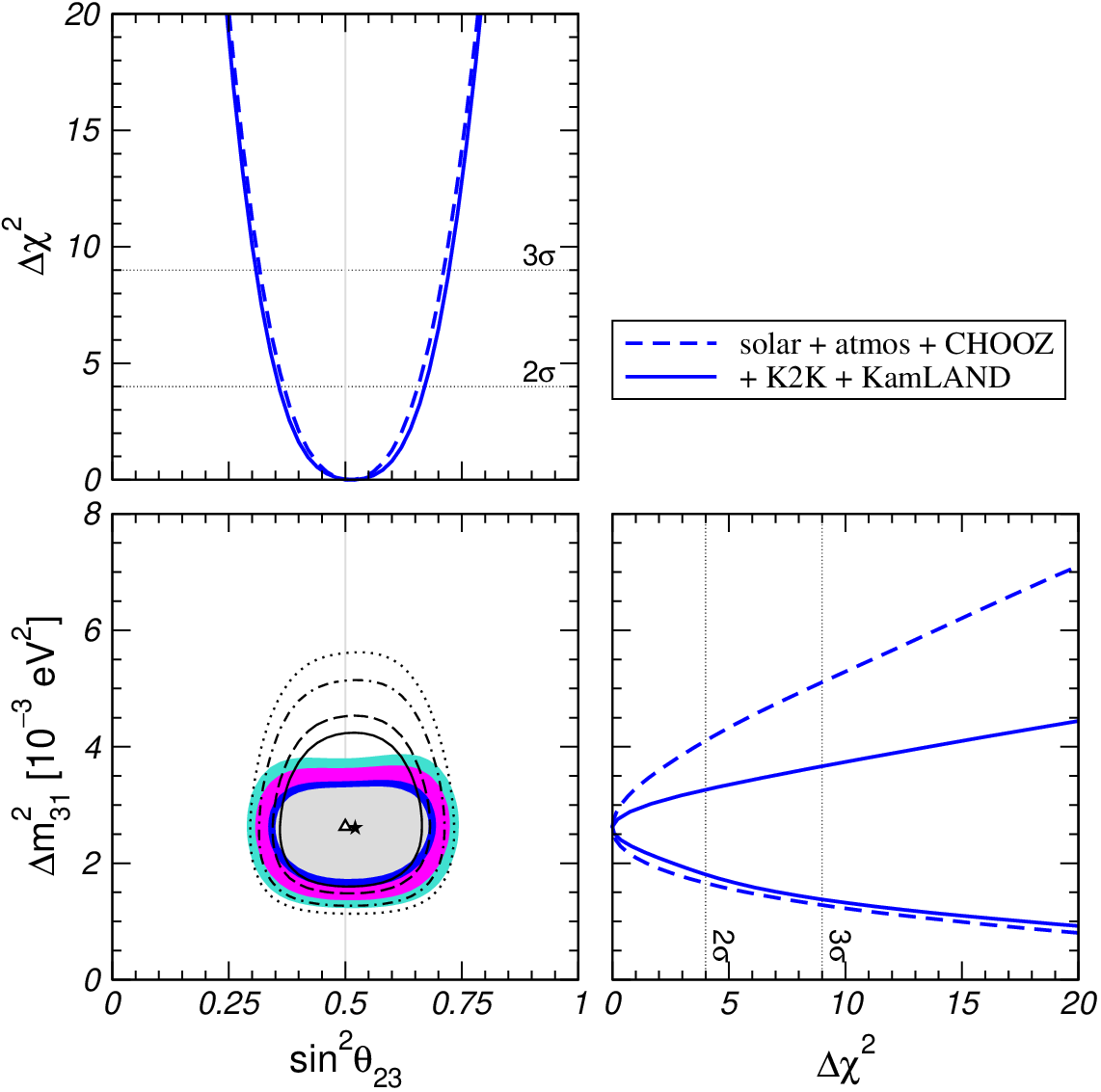}
    \caption{\label{fig:atm+k2k} 
      Two-neutrino atmospheric neutrino oscillation parameters from
      Ref.~\protect\cite{maltoni:2003da}. }  \vglue-.5cm
\end{figure}

Turning to the atmospheric neutrino parameters, we show in
Fig.~\ref{fig:atm+k2k} the projection of the allowed regions from the
global fit of all atmospheric data (details in
Ref.~\cite{maltoni:2003da})~\cite{ahn:2002up}, onto the plane of the
atmospheric neutrino parameters.  The regions displayed correspond to
90\%, 95\%, 99\%, and 3$\sigma$ \CL\ for 2 \dof\ implied by the
atmospheric+solar+CHOOZ data, while for the shaded regions also the
K2K and KamLAND data are added.  Also shown is the $\Delta \chi^2$ as
a function of $\sin^2\theta_\Atm$ and $\Dma$, minimized with respect
to undisplayed parameters.  One sees that the first 29 events from
K2K~\cite{ahn:2002up} included here already constrain the upper region
of $\Dma$.  This should be contrasted with the lowering of $\Dma$
indicated by a recent preliminary reanalysis of the atmospheric data
by the Super--Kamiokande collaboration presented at the Aachen EPS
conference~\cite{hayato:2003}.  While the two analyses differ, the
value for $\Dma$ quoted in \cite{hayato:2003} is statistically
compatible with the result shown above.  For $\Dma = 2\times
10^{-3}~\eVq$ and maximal mixing Ref.~\cite{maltoni:2003da} obtains a
$\Delta \chi^2 = 1.3$.

\section{Three--Neutrino Parameters}
\label{sec:2}

\begin{figure}[h] \centering
\includegraphics[width=.9\linewidth,height=6cm]{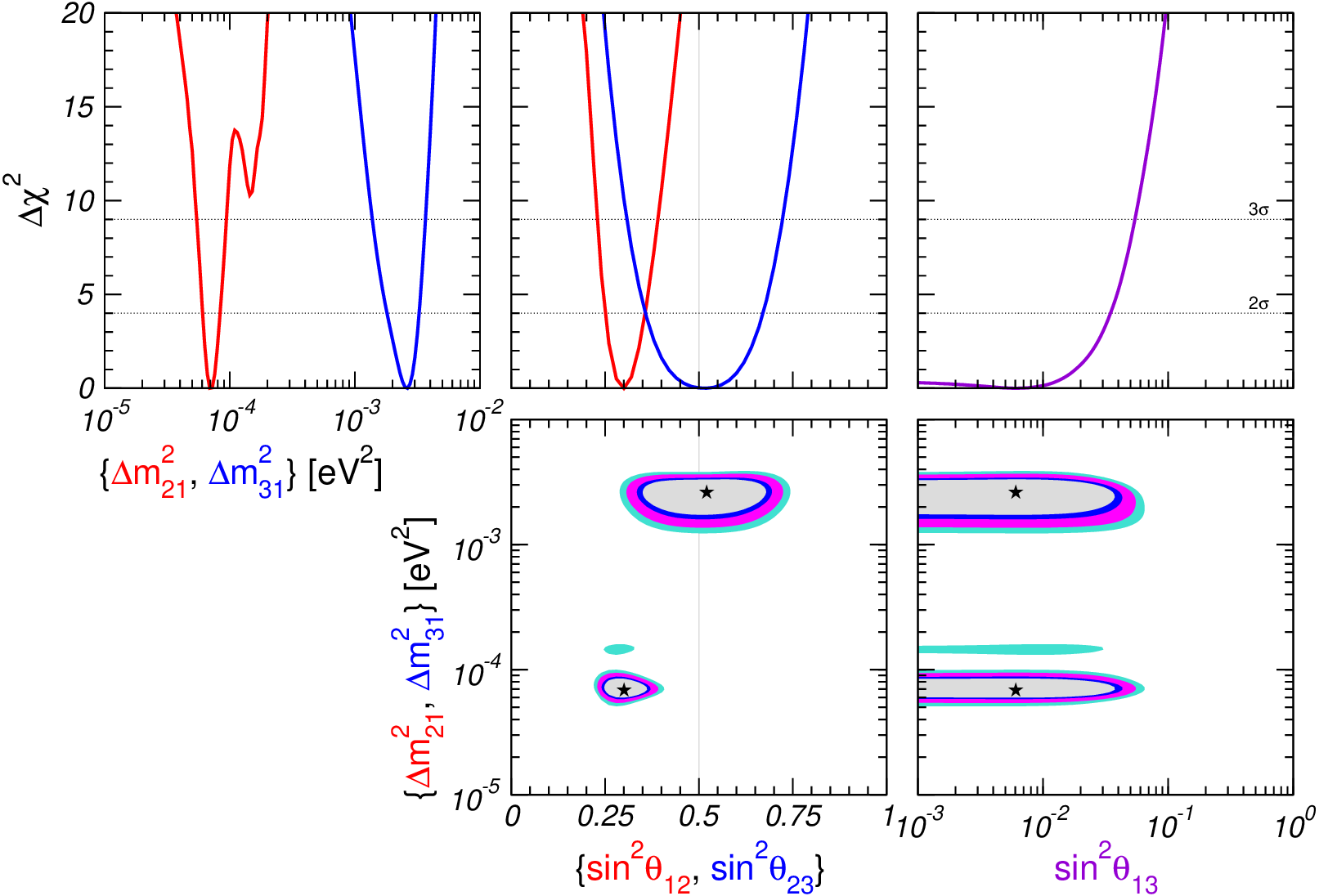}
    \caption{\label{fig:global} Three--neutrino
      oscillation parameters from Ref.~\protect\cite{maltoni:2003da}.}
\vglue-.5cm
\end{figure}

We now summarize the results of a global analysis combining all
current solar, atmospheric, reactor and accelerator data in order to
obtain the allowed three-neutrino oscillation
parameters~\cite{maltoni:2003da}.
The simpest three--neutrino lepton mixing matrix is parameterized as a
product of three complex rotations $K = \omega_{12} \omega_{13}
\omega_{23}$, $\omega_{ij}$ being a rotation in the $ij$ sector.  This
involves three mixing angles and three CP-violating
phases~\cite{schechter:1980gr}, one of which is the analogue of the
quark CP phase, whose effect in oscillations we neglect, while the two
Majorana phases~\cite{schechter:1980gr} do not show up in oscillations
but appear in lepton number violating
processes~\cite{schechter:1981gk,doi:1981yb}.  This way one is left
with just the three angles in the neutrino oscillation analysis:
$\theta_{12} \equiv \theta_\Sol$ which governs solar neutrino
oscillations, $\theta_{23} \equiv \theta_\Atm$ which characterizes
atmospheric neutrino oscillations, and $\theta_{13}$ which couples
these two analyses.
\begin{equation} \label{eq:mixing}
    K = 
\left(
    \begin{array}{ccc}
        c_{13} c_{12}
        & s_{12} c_{13}
        & s_{13} \\
        -s_{12} c_{23} - s_{23} s_{13} c_{12}
        & c_{23} c_{12} - s_{23} s_{13} s_{12}
        & s_{23} c_{13} \\
        s_{23} s_{12} - s_{13} c_{23} c_{12}
        & -s_{23} c_{12} - s_{13} s_{12} c_{23}
        & c_{23} c_{13}
    \end{array} \right) \,,
\end{equation}
Oscillations also involve the neutrino mass-squared differences $\Dms
\equiv \Delta m^2_{21} \equiv m^2_2 - m^2_1$ and $\Dma \equiv \Delta
m^2_{31} \equiv m^2_3 - m^2_1$.  Because of the hierarchy $\Dms \ll
\Dma$ it is a good approximation to set $\Dms = 0$ in the analysis of
atmospheric and K2K data, and to set $\Dma$ to infinity for the
analysis of solar and KamLAND data. The global fit to the data then
involves five oscillation parameters $\sin^2\theta_{12},
\sin^2\theta_{23}, \sin^2\theta_{13}, \Delta m^2_{21}, \Delta
m^2_{31}$.
The results of such three--neutrino analysis are summarized in
Fig.~\ref{fig:global}, taken from Ref.~\cite{maltoni:2003da}, showing
the allowed regions and $\chi^2$ projections for the above five
oscillation parameters.
The regions are at 90\%, 95\%, 99\%, and 3$\sigma$ \CL\ for 2 \dof\ 
for various parameter combinations. Also shown is $\Delta \chi^2$ as a
function of the five oscillation parameters, minimized with respect to
all undisplayed parameters.  Finally, the best-fit values, 2$\sigma$,
3$\sigma$ and 5$\sigma$ intervals (1 \dof) for the three-flavour
neutrino oscillation parameters derived from current solar,
atmospheric, reactor (KamLAND and CHOOZ) and accelerator (K2K)
experiments are given in Table 1 of Ref.~\cite{maltoni:2003da}. It is
remarkable that, for the first time, one can determine solar neutrino
parameters at the 5$\sigma$ level, showing that neutrino physics has
now entered the precision phase.

\section{Future agenda}
\label{sec:3}

\begin{figure}[h]
\begin{center}
\includegraphics[height=4.5cm,width=4cm]{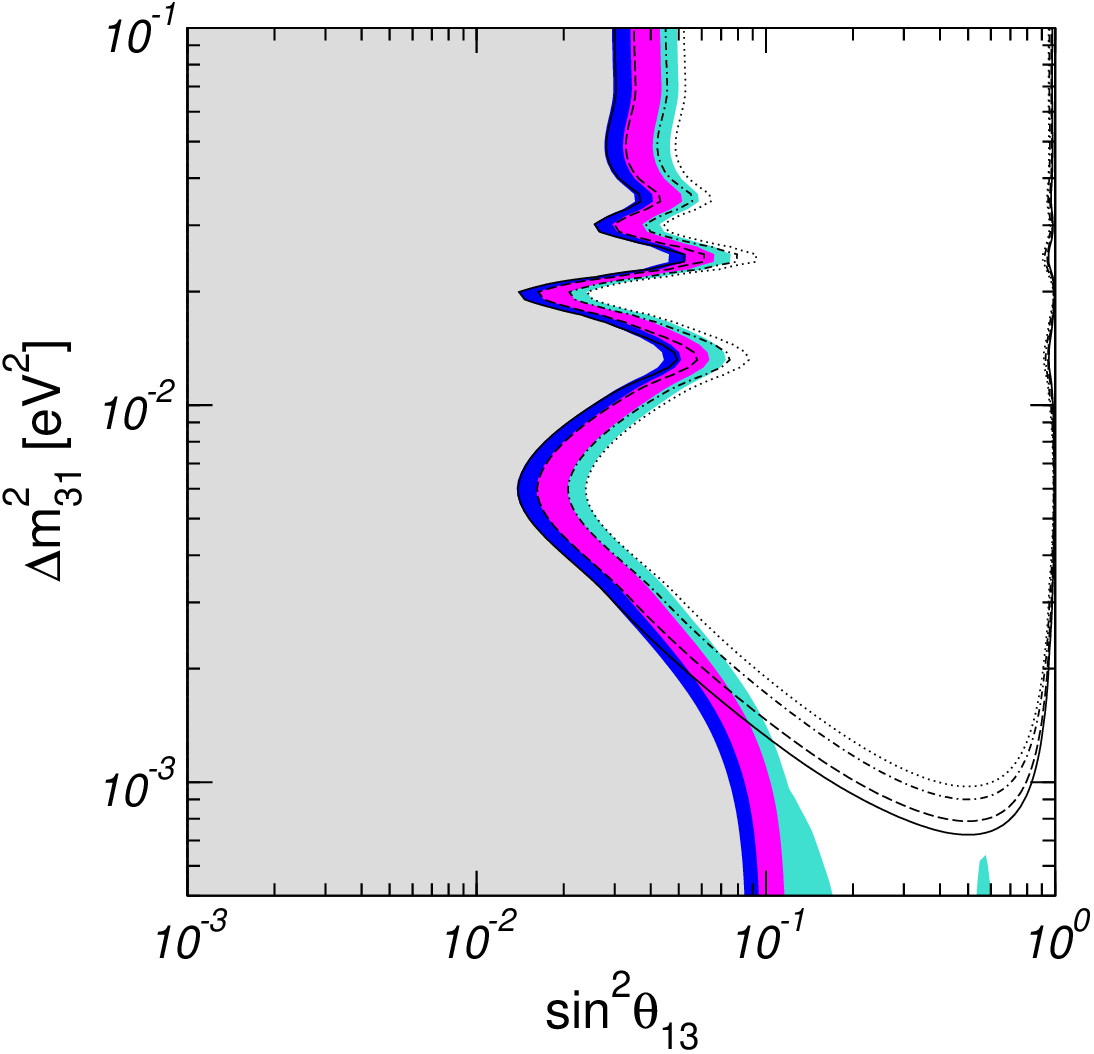}
\includegraphics[height=4.5cm,width=4cm]{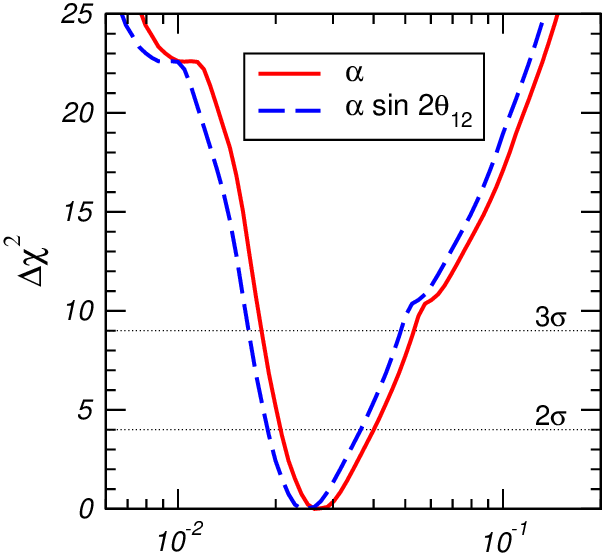}
\end{center}
\caption{\label{fig:t13-solar-chooz}%
   $\sin^2\theta_{13}$ and  $\alpha \equiv \Dms
  / \Dma$ from current neutrino data, from
  Ref.~\protect\cite{maltoni:2003da}.}
\end{figure}

So far all CP phases are neglected in all current neutrino oscillation
analyses. This is justified because the CP violating effects are
suppressed due to the stringent limits on $\theta_{13}$ following
mainly from reactor data~\cite{apollonio:1999ae}, shown in
Fig.~\ref{fig:t13-solar-chooz}.  On the left panel one can see the
90\%, 95\%, 99\%, and 3$\sigma$ allowed ($\sin^2\theta_{13}, \Dma$)
regions from CHOOZ data alone (lines) and CHOOZ+solar+KamLAND data
(shaded regions).  Moreover leptonic CP violating effects are
suppressed by the small mass splitting indicated by the solar neutrino
data analysis.  Indeed, in the 3-neutrino limit, CP violation
disappears as two neutrinos become degenerate~\cite{schechter:1980bn}.
Current data determine the ratio $\alpha \equiv \Dms / \Dma$ as shown
in the right panel of Fig.~\ref{fig:t13-solar-chooz}.

Now that the neutrino oscillation phenomenon has been confirmed, one
may try to go a step further and test for the phenomenon of leptonic
CP violation, either induced by the Dirac phase (oscillations) or by
the Majorana
phases~\cite{schechter:1980gr,doi:1981yb,schechter:1981gk} through
L-and-CP violating processes~\footnote{ Depending on the model,
  leptogenesis may involve both Dirac and Majorana
  phases~\cite{fukugita:1986hr}.}.
Let us start with oscillations.
One sees that the value for $\alpha$ inferred from the global neutrino
oscillation analysis and the reactor bound on $\sin^2\theta_{13}$ both
limit the prospects for probing CP violating effects at future
neutrino oscillation experiments with superbeams or neutrino
factories~\cite{apollonio:2002en,freund:2001ui,albright:2000xi,cervera:2000kp}.
It will be challenge to probe such small effects, and this will
require a near--detector in order to reject against the presence of
non--standard neutrino interactions~\cite{huber:2001de,huber:2002bi}.

On the other hand, now that neutrino masses have been established, it
is natural to check whether neutrinos are Majorana particles, as
expected from theory~\cite{schechter:1980gr}.  Neutrinoless double
beta decay~\cite{wolfenstein:1981rk} provides the most sensitive probe
into the nature of neutrinos, irrespective of its theortetical origin
~\cite{schechter:1982bd}.
There is indeed a new generation of proposed experiments aimed at
detecting \nbb with improved
sensitivity~\cite{elliott:2002xe,klapdor-kleingrothaus:1999hk}.

Although potentially sensitive to the Majorana CP phases present in
the lepton mixing matrix~\cite{schechter:1980gr,schechter:1981gk},
current nuclear physics uncertainties still preclude a realistic way
to test Majorana phases using this process~\cite{Barger:2002vy}, even
if several isotopes are combined.  As for other lepton number
violating processes, these are strongly suppressed by the small masses
of neutrinos and/or the V-A nature of the weak interaction. For
example the L-violating neutrino oscillation probability involved in
the ``thought-experiment'' proposed in Ref.~\cite{schechter:1981gk} is
suppressed by $(m_\nu/E)^2$, while transition Majorana neutrino
magnetic moments~\cite{schechter:1981hw} also vanish in the massless
neutrino limit~\cite{pal:1982rm}.

Let us now turn to another issue, namely, the number of light
neutrinos.  Are there more than three light neutrinos?

The LSND collaboration has claimed evidence for
oscillations~\cite{aguilar:2001ty} which would strongly suggest the
existence of a fourth (singlet) neutrino species at the electron-volt
range, as could arise, say, due to some protecting global symmetry
such as lepton
number~\cite{peltoniemi:1993ec,caldwell:1993kn,giunti:2000vv}.
However, a combined global four--neutrino study including also the
solar, atmospheric and negative short--baseline oscillation searches,
such as Karmen, Bugey and CDHS, strongly prefer the minimal three
light--neutrino
hypothesis~\cite{maltoni:2002ni,maltoni:2002xd,maltoni:2003yr}.
The data rule out the possibility of symmetric (2+2) schemes, because
in this case sterile neutrinos take part in both solar and atmospheric
oscillations. Though strongly disfavoured by short-baseline
experiments, the presence of a light sterile neutrino in a (3+1)
scheme may still be allowed, since it can be chosen to decouple from
solar and atmospheric oscillations.  Data from cosmology, including
CMB data from WMAP~\cite{spergel:2003cb}
\cite{crotty:2003th,elgaroy:2003yh,hannestad:2003xv} and the 2dFGRS
large scale structure surveys~\cite{tegmark:2001jh} lead to further
restrictions, especially on large $\Dml$ values. 

\section{Neutrino Theory: Top-Down versus Bottom-Up}
\label{sec:2}

The theoretical setting involved in the description of current
neutrino oscillation experiments was laid out long
ago~\cite{schechter:1980gr}, including the two-component quantum
description of massive Majorana neutrinos and the gauge theoretic
characterization of the lepton mixing matrix.  The other crucial
ingredient was the formulation of neutrino oscillations in the
presence of matter~\cite{wolfenstein:1978ue,mikheev:1985gs}.

The origin of neutrino mass remains as much of a mystery today as it
was back in the eighties.  Much of the early theoretical effort was
motivated in part by the idea of unification which introduced the
seesaw mechanism~\cite{gell-mann:1980vs,yanagida:1979}.  Although
first formulated in the context of the \10 group, it was soon realized
that the seesaw idea can be applied to left-right symmetric
theories~\cite{mohapatra:1981yp}, or the simplest effective Standard
Model gauge
framework~\cite{schechter:1980gr,chikashige:1981ui,schechter:1982cv}.
While the \10 or \lr seesaw formulations have the virtue of relating
the small neutrino mass to the dynamics of parity (gauged B-L)
violation, the effective \21 description is more general and applies
to \texttt{any} theory, for example with ungauged
B-L~\cite{chikashige:1981ui,schechter:1982cv}.  It is also
worth--noting that the general seesaw scheme implies a Higgs triplet
contribution to neutrino masses, from an induced tadpole or an
elementary scalar vacuum expectation
value~\cite{schechter:1980gr,mohapatra:1981yp,cheng:1980qt}.

However, it is worth stressing that the seesaw is just one way of
generating the fundamental dimension--five neutrino mass
operator~\cite{weinberg:1980mh}. Such may also arise from physics
``just around the corner''.  One example is provided by certain
super-string-inspired models~\cite{mohapatra:1986bd}. Indeed in
such ``anti-seesaw'' models neutrino masses vanish as the B-L scale
goes to zero, rather than infinity.

An alternative origin for neutrino mass is provided by the idea of low
energy supersymmetry~\cite{diaz:2003as,hirsch:2000ef,romao:1999up} in
schemes that break R parity through a sneutrino vacuum expectation
value~\cite{ross:1985yg,masiero:1990uj}. These lead effectively to
bilinear R parity violation~\cite{diaz:1998xc}.  The novelty here is
that neutrino mixing angles can be tested at accelerator
experiments~\cite{hirsch:2002ys,porod:2000hv,restrepo:2001me}.  Hybrid
alternatives involving triplet Higgs bosons and supersymmetry are
possible~\cite{aristizabal:2003ix}.

In summary there is no ``road--map'' for the ultimate theory of
neutrino mass, a wide variety of pathways remains open. In this context
one expects small residual effects associated to non-standard weak
interaction properties of neutrinos. These may follow from the
particular structure of the charged and neutral currents expected in
theories where neutrino masses follow from the existence of isosinglet
leptons~\cite{schechter:1980gr} or from alternative low energy
radiative mechanisms for neutrino mass generation~\cite{babu:1988ki}
and their variants.

\vspace{5pt}

\noindent
This work was supported by Spanish grant BFM2002-00345, European
Commission RTN grant HPRN-CT-2000-00148, European Science Foundation
Neutrino Astrophysics Network and a Humboldt Research Award. This
written version includes the SNO--salt data, analysed in detail in
Ref.  \cite{maltoni:2003da}.

\end{document}